# Understanding Disorder in Monolayer Graphene Devices with Gate-Defined Superlattices

**Vinay Kammarchedu,**[1,2,a,b] **Derrick Butler,**[1,2,4,a] **Asmaul Smitha Rashid,**[1,2] **Aida Ebrahimi,**[1,2,3,b] **and Morteza Kayyalha**[1,2,b]

[1]Department of Electrical Engineering, The Pennsylvania State University, University Park, Pennsylvania 16802, United States of America
[2]Materials Research Institute, The Pennsylvania State University, University Park, Pennsylvania 16802, United States of America
[3]Department of Biomedical Engineering, The Pennsylvania State University, University Park, Pennsylvania 16802, United States of America
[4]Current address: National Institute of Standards and Technology, Gaithersburg, MD 20899, United States of America
a) V. Kammarchedu and D. Butler contributed equally to this work.
b) Authors to whom correspondence should be addressed: vinay@psu.edu, sue66@psu.edu, mzk463@psu.edu

## Abstract

Engineering superlattices (SLs) – which are spatially periodic potential landscapes for electrons – is an emerging approach for the realization of exotic properties, including superconductivity and correlated insulators, in two-dimensional materials. While moiré SL engineering has been a popular approach, nanopatterning is an attractive alternative offering control over the pattern and wavelength of the SL. However, the disorder arising in the system due to imperfect nanopatterning is seldom studied. Here, by creating a square lattice of nanoholes in the $SiO_2$ dielectric layer using nanolithography, we study the superlattice potential and the disorder formed in hBN-graphene-hBN heterostructures. Specifically, we observe that while electrical transport shows distinct superlattice satellite peaks, the disorder of the device is significantly higher than graphene devices without any SL. We use finite-element simulations combined with a resistor network model to calculate the effects of this disorder on the transport properties of graphene. We consider three types of disorder: nanohole size variations, adjacent nanohole mergers, and nanohole vacancies. Comparing our experimental results with the model, we find that the disorder primarily originates from nanohole size variations rather than nanohole mergers in square SLs. We further confirm the validity of our model by comparing the results with quantum transport simulations. Our findings highlight the applicability of our simple framework to predict and engineer disorder in patterned SLs, specifically correlating variations in the resultant SL patterns to the observed disorder. Our combined experimental and theoretical results could serve as a valuable guide for optimizing nanofabrication processes to engineer disorder in nanopatterned SLs.

## 1. Introduction

The ability to engineer the potential landscape for electrons in two-dimensional (2D) materials is an emerging strategy to study exotic phases of matter [1], ranging from unconventional superconductors [2] and correlated insulators [2] to Wigner crystals [3] and Chern insulators [4–7]. The realization of these devices is possible due to the formation of a periodic potential by superlattice (SL) engineering. Several methods have been explored for SL engineering in 2D materials, including periodic heteroatom doping [8], pattern etching [9], moiré pattern (twist angle engineering or lattice mismatch) [10–18], strain engineering [19], ferroelectric domain gating [20], and patterned electrostatic gating [9,21–29]. Although moiré heterostructures have been widely popular, they face several reliability problems. One of the main issues is the disorder originated from inhomogeneous angle and strain [29]. Further complications due to domain formation and lattice relaxation also affect the reproducibility of these devices [30]. On the other hand, patterned electrostatic gating allows for user-defined geometry (e.g., triangular, square, etc.) and variable SL sizes. Furthermore, the SL formed using a periodic electrostatic field allows for the in-situ control and modulation of the SL potential [31].

SL engineering of hBN-encapsulated graphene devices via periodic electrostatic fields has been achieved by patterning a dielectric [23] between graphene and the gate [25], as well as using a patterned gate electrode with a uniform dielectric material [32,33]. In both cases, the periodic electric field can have a controlled SL pattern and wavelength. Furthermore, the strength of the SL potential can be tuned on demand [25]. However, SL engineering via patterning is mostly limited by the nanofabrication process [33]. One of the challenges in fabricating patterned SLs via nanolithography is the control of SL pattern disorder which often limits the visibility of the SL effects during measurements [34]. In experiments, state-of-the-art encapsulated graphene devices usually have a residual carrier concentration ($n_0'$) in the range of ~ $10^{10}$ cm$^{-2}$ [35]. This sets the disorder energy scale at about $\hbar v_F \sqrt{\pi n_0'}$ ~ 10 meV, where $\hbar$ and $v_F$ are the reduced Planck constant and the Fermi velocity, respectively [34]. However, due to variations in device fabrication processes, residual concentrations on the order of $10^{11}$ to $2.5 \times 10^{11}$ cm$^{-2}$ are common [36,37], resulting in a disorder energy as large as 50 meV [34]. The SL effects have been predicted to emerge at such a large disorder level, albeit for nanopatterns with dimensions of a few tens of nanometers [29]. This size limitation is approaching the limit of what is possible using nanofabrication methods such as electron beam lithography (EBL) [33]. Furthermore, lithography itself could introduce variations in the device structure and add to the disorder of the system [34,37–39]. It is, therefore, imperative to study and understand disorder effects in nanofabricated SLs to curate highly efficient fabrication processes. Previous studies have considered the variations produced by lithography in device structures that directly etch the graphene channel [40–43]. However, the impact of disorder is yet to be investigated in graphene heterostructures with nanopatterned dielectric substrates.

In this work, we study SL effects and disorder in hBN-encapsulated monolayer graphene [36,44] devices with a graphite top gate and a patterned SiO$_2$ dielectric layer. To investigate the SL effects, we perform low-temperature electrical transport and Hall measurements and observe that by increasing the SL potential, satellite SL peaks appear in the resistance versus carrier concentration data. In addition, our data show that the disorder in the patterned SiO$_2$ graphene is an order of magnitude higher than that in unpatterned SiO$_2$ graphene devices. To elucidate the underlying sources of the disorder, we perform finite-element modeling of the electric field. We specifically consider nanohole disorder in the form of nanohole size variations, adjacent nanohole mergers, and nanohole vacancies. Using a resistor network model, we characterize the impact of the disorder on the electrical resistance of graphene. We show that among the various sources, the variation in the nanohole size is the dominant factor in our devices. More specifically, we show that the full-width half maximum (FWHM) of the resistance peak changes by 600% for 5% nanohole size disorder, whereas the FWHM changes by 700% for 5% nanohole size variations and 3% adjacent nanohole mergers. We correlate the theoretically calculated disorder to our experimental results using topography characterization of the SL and find that the increase in disorder of our graphene devices closely matches the theoretical prediction. We also utilize quantum transport simulations coupled with the finite-element modeling to qualitatively confirm the results. This study hence provides a framework to predict and engineer disorder in patterned SLs, specifically correlating variations in the resultant SL patterns to the observed disorder.

## 2. Materials and Methods

### 2.1 Superlattice fabrication

The SL is prepared on a 1 cm$^2$ Si/SiO$_2$ wafer (285 nm dry, thermal oxide; NOVA Wafer) with Cr/Au alignment markers (5 nm/ 45 nm). The wafers are dehydrated for 20 mins at 180°C before spin coating a mixture of ZEP520A and anisole (1:1 ratio) at 6000 rpm for 45 s. The samples are then baked on a hot plate for 3 mins at 180°C. The SL is patterned using electron-beam lithography (Raith EBPG5200) with a beam current of 500 pA, beam step size of 5 nm, and a dose of 400 – 435 µC cm$^-$

². The pattern is designed to be a square SL (area, $A^2 = 40^2$ nm²) formed by circular nanoholes (radius, r = 12.5 nm) with an etch depth of ~ 30 nm. The patterns are developed in n-Amyl acetate at -10°C for 3 mins followed by rinsing in isopropanol (IPA) for 1 min and drying with $N_2$. Dry etching is carried out in a Plasma-Therm Versalock 700 inductively coupled plasma system. The chamber is cleaned for 15 mins under an $O_2$ environment at 800 W (ICP power). The $SiO_2$ is then etched for 30 s at 5 mTorr in a mixture of $CHF_3$ (30 sccm) and $CF_4$ (10 sccm) with a chuck power of 50 W and ICP power of 300 W. The depth of the etched regions is ~ 30 nm. After etching, the resist is removed in dimethyl sulfoxide (DMSO) heated to ~ 80 °C for at least 1 h followed by rinsing in IPA and DI water. To remove any residual resist, the samples are exposed to an $O_2$ plasma (Harrick Plasma) for 10 mins at 30 W and 840 mTorr. After plasma cleaning, the samples are further cleaned in Nanostrip heated to 60°C for 20 mins and thoroughly rinsed with DI water before drying.

## 2.2 Heterostructure assembly

Monolayer graphene (Kish Graphite, CoorsTek Inc.) is mechanically exfoliated and identified via optical contrast. A top graphite is also exfoliated and picked up first followed by the top hBN (15 nm, HQ Graphene), the graphene, and bottom hBN (5 nm). Standard electron beam lithography is used to pattern and etch the heterostructure. Another EBL step is performed to define the Cr (10 nm)/Au (100 nm) contacts.

## 3. Results and Discussion

### 3.1 Topographic analysis of the nanopatterned superlattice

Figures 1a and b show the schematic of our nanopatterned SL graphene heterostructure and an optical image of the device, respectively. Figure 1b also depicts an atomic force microscope (AFM) image of the nanopatterned SL. We fabricate a graphite/hBN/graphene/hBN heterostructure on a 285 nm $SiO_2$/Si substrate using a standard dry transfer technique. Prior to the heterostructure transfer, we nanopattern the $SiO_2$ dielectric into a square-patterned SL, consisting of nanoholes with a diameter of approximately 25 nm and pitch size of approximately 40 nm. We encapsulate graphene with a 5-nm-thick bottom hBN flake which is thin enough to preserve the spatial SL electric field pattern; more information about the device fabrication can be found in the supplementary information (SI). As shown in the AFM image of Fig. 1b, we observe a uniform square pattern over a 4 μm² area. We identify two types of defects in the SL from this AFM image. First, using image analysis of the AFM data, we estimate ~ 5% variations in the nanoholes size (represented by $\gamma_r$); see SI. For this estimation, we assume a normal distribution for the radius of nanoholes with $12.5 \times \frac{\gamma_r}{100}$ nm as the standard deviation and 12.5 nm as the mean. Second, we observe ~ 3% nanohole mergers in the SL when two or more nanoholes have merged. We represent this by $\gamma_m$, which is the percentage of the number of mergers per number of nanoholes. We also consider nanohole vacancies where the nanofabrication failed to yield a nanohole. We represent this by $\gamma_v$, which is the percentage of the number of vacancies per number of nanoholes. We note, however, that our image analysis does not recognize any vacancies in our patterns. Therefore, we will primarily use $\gamma_r$ and $\gamma_m$ to compare our experimental results with the theoretical model.

### 3.2 Electrical transport measurements

We perform electrical transport measurements using standard lock-in amplifiers (SR860, Stanford Research Systems) in a Bluefors dilution refrigerator (LD250) with a base temperature of 10 mK. All measurements are performed at 10 mK unless noted otherwise. Figure 1c plots the longitudinal resistance ($R_{xx}$) for various carrier concentrations and back-gate voltages. We calculate the carrier concentration ($n$) using a parallel-plate capacitor model. Specifically, we use the trajectory of the main Dirac point as a function of back- and top-gate voltages to calculate the intrinsic carrier density $n_0'$ using a double capacitor model as $n = C_{tg}V_{tg} + C_{bg}V_{bg} + n_0'$ [25,37]. By fitting the charge neutrality point ($n = 0$), we obtain $n_0' \approx 4.3 \times 10^{11}$ cm⁻²; see Section. S3 of the SI for more details. Figure 1c shows the expected main Dirac peak near the charge neutrality point. We further observe mini peaks around the charge neutrality point (blue arrows in Fig. 1c), which only emerge when the SL potential is large ($V_{bg} > 40\ V$). This observation is consistent with previous reports [25], that the satellite peaks are due to the spatially periodic electric field formed by the nanopatterned dielectric. For a square SL, the first satellite peak is predicted to appear at $4n_0 = \frac{4}{A^2} = 2.5 \times 10^{11}$ cm⁻² where $A^2$ is the area of the SL unit cell ($A \approx 40$ nm) and $n_0 = \frac{1}{A^2}$ [34]. In our device, the first satellite peak occurs at $n = -2.5 \times 10^{11}$ cm⁻² which is consistent with the predicted $-4n_0$ for a square SL [25]. The second peak, however, appears at $n = -12n_0 = -7.5 \times 10^{11}$ cm⁻² which is larger than the expected peak around $|n| = 6.5n_0$ for a square lattice. We note that this disparity has also been observed in previous reports [25], and is attributed to the simplicity of the model which underestimates the lattice strength and hence overestimates the bandgap overlap in the experimental system.

Because of this disparity, we perform quantum transport simulations of a scaled graphene device on a square SL (see SI). We find theoretically that the two main SL peaks appear at $n = -4n_0$ and $n = -12n_0$, respectively, corroborating our experimental observations. We also perform magneto transport measurements to confirm SL effects as described in the SI. Finally, we note that the observed satellite peaks are weaker on the electron side with only one clear peak. This non-symmetry has previously been observed [25] and is associated to the electron-hole asymmetry of the band structure, resulting in an asymmetry in the resistance vs carrier concentration curves.

To analyse the disorder, we consider the FWHM of the resistance versus carrier density data. We find that the peaks in Fig. 1c are broad (FWHM ~ $2.4 \times 10^{11}$ cm$^{-2}$) indicating a high disorder level in our graphene heterostructure. However, this FWHM is comparable to previously reported data for similar devices e.g., FWHM $\approx 2 \times 10^{11}$ cm$^{-2}$ for 35 nm square SL and FWHM $\approx 3.5 \times 10^{11}$ cm$^{-2}$ for 40 nm triangular SL [25]. We note that the FWHM is commonly used to analyze disorder in data with a single peak [35]. Since our data contains multiple overlapping peaks, we use a Gaussian distribution for the constituent peaks and fit them to the resistance vs. carrier concentration data shown in Fig. 1c for $V_{bg}$ = 70 V. The fitted curve closely matches the raw data as seen in Fig. 1d. From these constituent peaks, we individually extract the FWHMs as $2.4 \times 10^{11}$ cm$^{-2}$ and $2.8 \times 10^{11}$ cm$^{-2}$ for the main peak and the satellite peaks, respectively. This FWHM is an order of magnitude larger than that of graphene devices on non-patterned substrates [35]. This observation serves as a motivation for us to further investigate the sources of disorder in dielectric-patterned SL devices in the later part of the manuscript.

### 3.3 Finite-element modeling

We perform finite-element modeling using COMSOL Multiphysics version 6.2 and MATLAB version R2023a to better understand how nanolithography limitations in fabricating nanopatterned dielectric SLs affects the electrostatic gating across the sample. We use the Electrostatics module with a physics-based mesh specified to parameter "3". The model comprises a Si back gate ($V_{bg}$) as a boundary condition placed at the base of a SiO$_2$ dielectric layer (285 nm and dielectric constant ε = 3.9). The SL is composed of nanoholes which are treated as ideal vacuum (ε = 1) with a depth of 30 nm, a radius ($r$) of 12.5 nm, and a pitch ($a$) of 40 nm. Figure 2a shows the simulation geometry, which comprises a 20 × 20 array of uniform nanoholes (radius $r$ = 12.5 nm, pitch size $a$ = 40 nm, depth = 30 nm) etched into a 285 nm thick SiO$_2$ volume. We include a 5 nm thick hBN layer (ε = 3) on top of the SiO$_2$ is and set the boundary condition to ground on top. In this figure, the hBN layer is excluded for clarity. Using a geometric capacitance model, we calculate the carrier concentration ($n$) and differential carrier concentration $\Delta n = n - \langle n \rangle$, where $\langle n \rangle$ is the average carrier concentration across the substrate. We use $\Delta n$ instead of $n$ in our calculation to highlight the variations in the carrier concentration. We allow the radii of the nanoholes to vary according to a Gaussian distribution centred around the prescribed radius of 12.5 nm and standard deviation of $12.5 \times \frac{\gamma_r}{100}$ nm, where $\gamma_r$ captures nanohole size variations. Moreover, we allow adjacent nanoholes to merge with a percent $\gamma_m$ and consider nanohole vacancies by including missing nanoholes with a percent $\gamma_v$. Figure 2b shows uniform distribution of $\Delta n$ for $\gamma_r$, $\gamma_v$, and $\gamma_m$ = 0%, i.e., a pristine SL. On the other hand, Figs. 2c-f show the distribution of $\Delta n$ for different combinations of $\gamma_r$, $\gamma_v$, and $\gamma_m$. Such variabilities in $\gamma_r$, $\gamma_v$, and $\gamma_m$ could arise in practice from minor instabilities and fluctuations during nanolithography. In this case, spatial inhomogeneity of the displacement field caused by non-zero $\gamma_r$, $\gamma_v$, and $\gamma_m$ leads to nonuniformity in the induced carrier concentration of graphene. To better quantify $\gamma_r$, $\gamma_v$, and $\gamma_m$ in our simulated SL, we calculate the average carrier density $n_{ij}$ for each of the unit-nanohole, i.e., a 40 nm by 40 nm square region as shown with dashed squares in Fig. 3a. This formulation allows us to look at the variation of the carrier density in each unit cell due to nanohole imperfections. Since our substrate has no nanohole vacancies, we focus on $\gamma_r$ and $\gamma_m$. We determine $\sigma_n$ which is the standard deviation of $n_{ij}$ for various $\gamma_r$ and $\gamma_m$ ($\gamma_v$ = 0%) and plot it in Fig. 3b using contour lines for visual clarity. We observe that $\sigma_n$ is close to zero for pristine SL ($\gamma_r$, $\gamma_v$, $\gamma_m$ = 0%) as expected. As $\gamma_m$ increases from 0 to 8%, we observe an increase of approximately $0.6 \times 10^{11}$ cm$^{-2}$ in $\sigma_n$, whereas, we observe a larger increase of approximately $1.1 \times 10^{11}$ cm$^{-2}$ when $\gamma_r$ increases from 0 to 8%. At low $\gamma_m$ < 3%, the effect of $\gamma_m$ and $\gamma_r$ is similar when compared independently. However, at higher levels ($\gamma_m$ > 3%), $\gamma_r$ is dominant. For $\gamma_r$ = 5% and $\gamma_m$ = 3% estimated from our AFM data, $\sigma_n$ is roughly $0.8 \times 10^{11}$ cm$^{-2}$. We will now discuss how this standard deviation ($\sigma_n$) is translated to an estimate for the FWHM.

### 3.4 Resistor network Simulation

To include the effects of nanohole disorder in the resistance, we model graphene using a resistor network, which has been successfully used to model charge puddles in graphene [45,46]. We simulate a square network of resistors as shown in Fig. 4a, wherein each node is connected to the right and bottom nodes via two similar resistors ($R_{ij}$). We determine $R_{ij}$ for each unit cell using the function $R(n = n_{ij}+n_{tg})$. The resistance $R$ is analytically modeled based on a response of a miniscule graphene device with FWHM of ~ $0.25 \times 10^{11}$ cm$^{-2}$ (see Section. S7 for analytical expression of $R$ and a plot of $R$ vs $n$ behaviour). In our system, the intrinsic disorder may be present due to the charge puddles at the interface of the encapsulated hBN-graphene-hBN heterostructure and the silicon dioxide substrate. The value of the FWHM depends on the substrate and can be changed based on the engineered device characteristics. For example, the FWHM may be higher when graphene is directly placed on SiO$_2$ substrate [36,47], or lower when graphite gates are used which screen charges at the SiO$_2$ interface [48]. To calculate the value of $R$, we plug in $n = n_{ij} + n_{tg}$. $n_{ij}$ is the average unit cell carrier density defined earlier and $n_{tg}$ is the carrier density induced by the top gate. This $n_{ij}$ depends on the back-gate voltage, and more importantly captures the variations in the SL patterns (see Fig. 3). We calculate the local carrier concentration $n_{ij}$ based on the finite-element simulation. $n_{tg}$ is the global carrier concentration induced by the top gate, which we assume is uniform throughout the substrate at a fixed top gate voltage. The local carrier concentration, however, $n_{ij}$ is in fact not uniform in a non-uniform pattern and can vary across the substrate based on nanopattern variations. We note that varying $\gamma_r$, $\gamma_v$, and $\gamma_m$ results in changes in $n_{ij}$ which will be captured by the function $R(n = n_{ij}+n_{tg})$. We calculate the resistance using the network of Fig. 4a and plot $R_{xx}$ versus $n$ in Fig. 4b. In order to calculate the $R_{xx}$ of the macroscopic device, $R_{ij}$ of each microscopic graphene device with varying $n_{ij}$ is calculated and simulated using a resistor network. The more variation present in the $n_{ij}$ more the $R_{xx}$ peak broadens due to misalignment of the charge neutrality points of the microscopic graphene devices. We observe that for $\gamma_r = 0$, full width half maximum is FWHM$_P = 0.35 \times 10^{11}$ cm$^{-2}$ which is slightly higher than the expected value of $0.25 \times 10^{11}$ cm$^{-2}$. We attribute this discrepancy to the numerical inaccuracy due to digitization of large carrier concentration numbers in the finite-element simulation.

*3.5 Quantum Transport Simulation and Outlook*

We observe that in Fig. 4b, with an increase in $\gamma_r$, the main Dirac peak broadens; this is because each SL unit cell has a slightly different carrier concentration ($n_{ij}$). As a result, the calculated FWHM increases with the increasing $\gamma_r$. Figure 4c plots the colormap of FWHM as a function of $\gamma_r$ and $\gamma_m$. Solid black lines with labels correspond to contours with fixed FWHM with varying $\gamma_r$ and $\gamma_m$. We observe that the constant FWHM contours are generally vertical. This indicates that $\gamma_r$ has a greater influence on the FWHM. We further observe that the FWHM increases by as high as 1000% for $\gamma_r$ and $\gamma_m$ around 8% when compared to FWHM$_P$. For our experimental conditions of $\gamma_r = 5\%$ and $\gamma_m = 3$ %, our model predicts the FWHM of ~ $2.4 \times 10^{11}$cm$^{-2}$ which is close to the experimentally observed FWHM (blue dot in Fig. 4c). In our device, $\gamma_r = 5\%$ alone corresponds to ~ 600% increase in FWHM compared to FWHM$_P$. When $\gamma_m$ is also increased from 0 to 3%, the FWHM increases by ~ 700%. This FWHM matches the one we estimate from our experimental results in Fig. 1c. To further confirm the results of our theoretical modeling, we perform quantum transport simulations where we use the output of the finite-element model as the SL potential input. In this case, we simulate a 0.5 µm$^2$ square area scaled graphene (scaling parameter = 4) lattice using a tight-binding model [49]. We perform all numerical calculations using a python-based Kwant code [50]. We consider two leads which are attached to the opposite sides of the square lattice. We calculate the SL potential from the carrier concentration obtained via the finite-element simulation. The SL potential nominally varies with a peak-peak value of 50 meV which is comparable to experiments for back-gated graphene devices with patterned dielectric [25,51,52]. To induce intrinsic disorder, we delete 0.01% of carbon atoms to form vacancies. We utilize a Savitzky–Golay filter (sampling distance of 0.1 meV, 35 window size, 2$^{nd}$ order) to smoothen the noisy data around the Dirac peak and to better estimate the relative FWHM. This noise is due to numerical variation of the SL potential from the finite-element simulation. Figure S5 of the SI shows the schematic of the quantum simulation and plots the calculated transmission as a function of carrier density and energy. By varying the disorder, especially $\gamma_r$ from 0 to 4%, we observe an increase in FWHM as the disorder increases; see Fig. S5c in the SI. Overall, we calculate a 3x increase in FWHM from FWHM$_P$ at $\gamma_r = 4\%$. This is comparable to our predicted value when FWHM$_P$ is around $10^{10}$ cm$^{-2}$.

Figure 4d plots the colormap of the FWHM as a function $\gamma_r$ and FWHM$_P$. We observe that the FWHM increases with increasing $\gamma_r$. Assuming the maximum tolerance of the FWHM to be around $4n_0$, we can estimate the maximum allowed FWHM$_P$ for each $\gamma_r$. For example, for a square SL of 40 nm wavelength, i.e., $4n_0 = 2.5 \times 10^{11}$ cm$^{-2}$, the maximum allowed $\gamma_r$ is around 6.5% for a high-quality graphene device with FWHM$_P \sim 10^{10}$ cm$^{-2}$. In fact, Fig. 4d can be used as a guide to estimate the expected disorder in the system by assuming a reasonable FWHM$_P$ and measuring the $\gamma_r$ prior to heterostructure fabrication. This will allow us to predict if the SL pattern will yield a device with low enough disorder. Therefore, assuming a state-of-the-art EBL limit of 30 nm in feature dimensions [33] and FWHM$_P = 10^{10}$ cm$^{-2}$, we estimate a maximum allowed $\gamma_r = 11\%$ to achieve the FWHM of $4.5 \times 10^{11}$ cm$^{-2} \leq 4n_0$. Our model, although described for a square SL of 40 nm wavelength, can easily be adapted to other SL geometries; see for example the results for a triangular lattice in section S8 of the SI. In contrast to square SL, we observe that $\sigma_n$ depends equally on both $\gamma_r$ and $\gamma_m$. This is because in a triangular lattice, each nanohole has six neighbouring nanoholes and a higher chance of nanohole fusion. Moreover, due to the smaller relative unit cell compared to a square lattice, each fusion leads to a higher susceptibility in the variation of the carrier concentration. Interestingly, we observe that triangular SL has higher resistance towards nanohole size variations as compared to the square lattice. This observation together with the fact that for the same wavelength, $4n_0$ in triangular SLs is much higher than square SLs make triangular SLs more robust against disorder. Although not a replacement for quantum simulations, we speculate our model is a simple and versatile tool to understand and evaluate disorder in nanopatterned SL devices.

## 4. Conclusion

In conclusion, we successfully demonstrated the use of a patterned dielectric superlattice to fabricate a device with in-situ tunable superlattice effects. We performed longitudinal resistance and Hall measurements to confirm the presence of the SL effects. We observed that the SL device had a higher disorder compared to unpatterned graphene devices. We investigated this disorder through a combination of modeling and experimental analysis. Specifically, we investigated three kinds of disorder: variations in the size of nanoholes, adjacent nanohole mergers, and nanohole vacancies. Furthermore, we modeled graphene using a resistor network to translate the variations in the simulated electrostatics to disorder in transport characteristics. We found that for square SLs, the disorder primarily originates from variations in the SL pattern formed during lithography. We further confirmed this finding using a quantum transport simulation model. The developed disorder model could offer new insights into ways to improve the superlattice quality in nanopatterned devices. Furthermore, beyond graphene, the developed model could be generalized to study other 2D materials and device structures for a variety of electronic applications. Our combined experimental and theoretical results could help determine the accepted disorder level prior to complex nanofabrication of 2D heterostructures with nanopatterned dielectric layers or gate electrodes.

## Acknowledgements

V.K., D.B., and A. E. would like to acknowledge partial support from the NSF:I/UCRC Center for Atomically Thin Multifunctional Coatings (ATOMIC) (Division Of Engineering Education and Centers, Award No. 2113864), NSF Division of Electrical, Communications and Cyber Systems (ECCS, Award No. 2236997), and the National Institutes of Health (Award No. R21EB031354). M.K. and A.S.R. acknowledge support from the Materials Research Science and Engineering Center (MRSEC) funded by the US National Science Foundation (DMR 2011839). The content of this report is solely the responsibility of the authors and does not necessarily represent the official views of the National Science Foundation (NSF) or National Institutes of Health (NIH). V.K. acknowledges the partial support by Penn State Leighton Riess Graduate Fellowship in Engineering. The authors would also like to thank the Roar supercomputing resources of the Penn State Institute for Computational and Data Sciences (ICDS).

**Conflicts of interest:** All authors declare that they have no conflicts of interest.
**Data availability:** The experimental and simulation data supporting the findings of this paper can be found in the Zenodo repository. DOI: doi.org/10.5281/zenodo.13221419

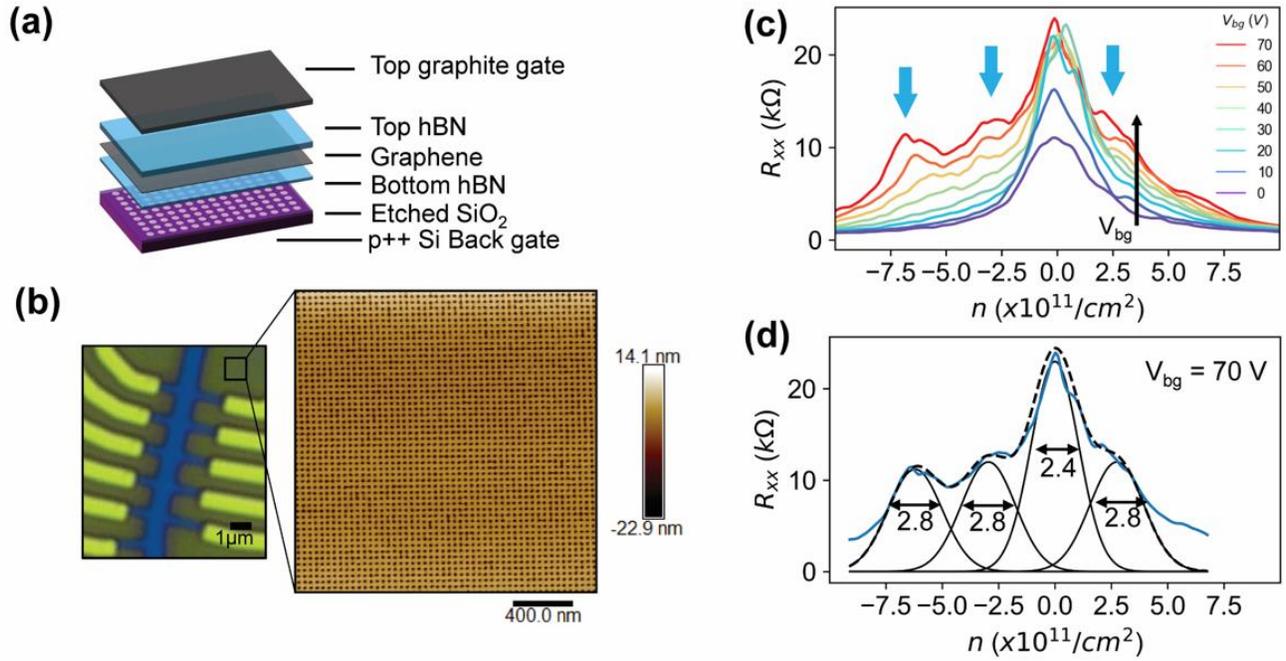

**Figure 1. Device structure and electrical transport at zero magnetic field.** (a) Schematic of the hBN-encapsulated heterostructure consisting of patterned SiO$_2$ as the back gate dielectric and a top graphite gate. (b) Optical image of the fabricated hall bar along with atomic force microscope (AFM) image of the patterned substrate. (c) Longitudinal resistance $R_{xx}$ as a function the carrier density $n$ at various back gate voltages ($V_{bg}$'s). The satellite peaks due to nanopatterned superlattice (SL) are marked by blue arrows. (d) The individual Gaussian peaks fitted to estimate the full width half maximum (FWHM) of each peak. The values of the FWHM are indicated within the panel for each peak.

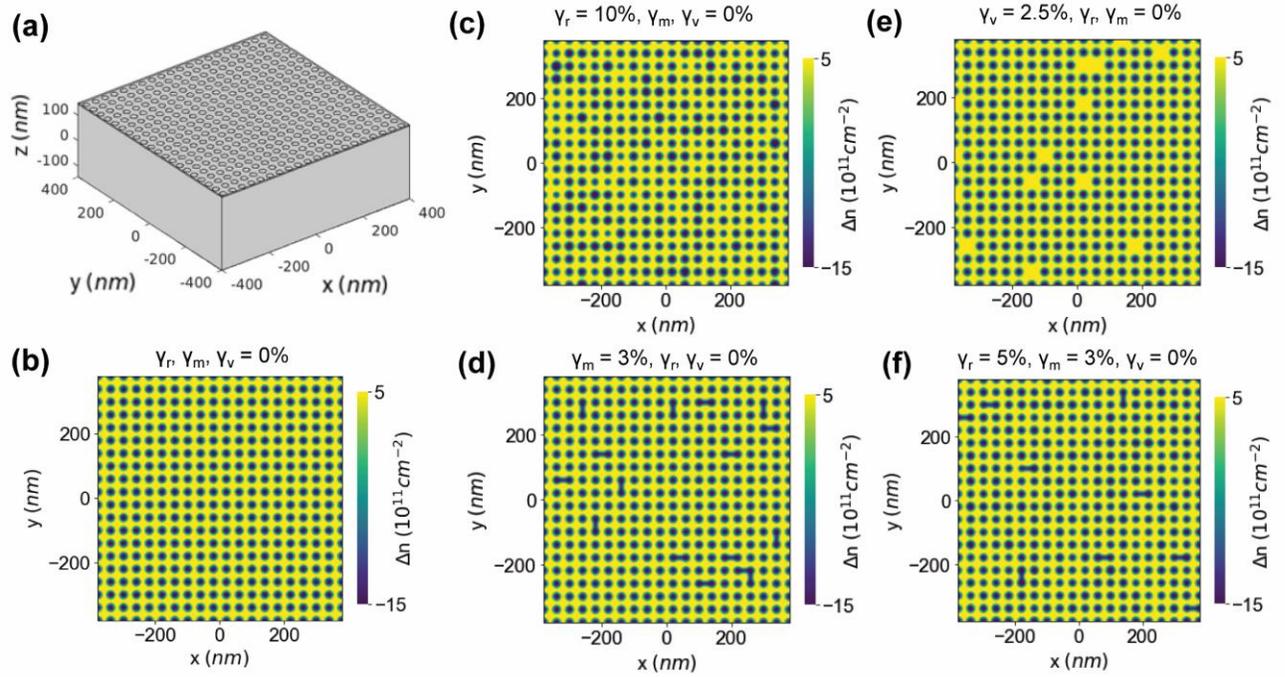

**Figure 2**. **Finite-element modeling**. (a) The geometry of a dielectric superlattice with constant radii used in the finite-element analysis. The top hBN layer is removed for clarity. (b) Exemplary maps of differential carrier density $\Delta n$ for a pristine superlattice (no variation in nanohole size $\gamma_r = 0\%$, no nanohole vacancy $\gamma_v = 0\%$, and no nanohole merger $\gamma_m = 0\%$) and a superlattice with randomly varying $\gamma_r$ (c), $\gamma_m$ (d), and $\gamma_v$ (e). (f) The map of $\Delta n$ corresponding to $\gamma_r = 5\%$, $\gamma_m = 3\%$, and $\gamma_v = 0\%$, corresponding to the topographically estimated variations from our AFM image.

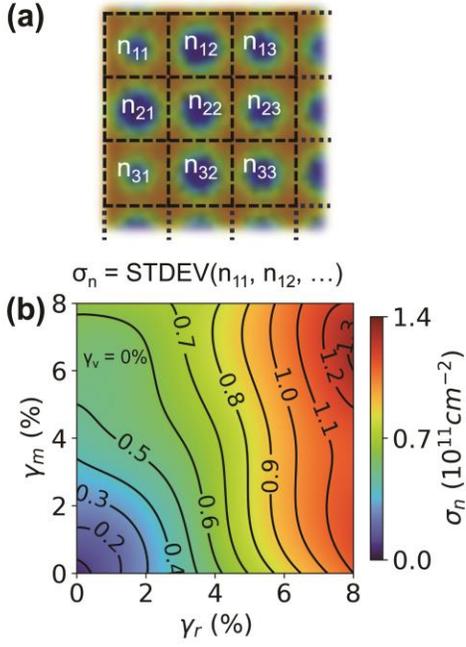

**Figure 3. Variations of the carrier concentration in the superlattice**. (a) The calculated average carrier density $n_{ij}$ for each unit-nanohole which is then used to calculate the standard deviation, $\sigma_n$. (b) The standard deviation $\sigma_n$ as a function of $\gamma_m$ and $\gamma_r$ with $\gamma_v = 0\%$.

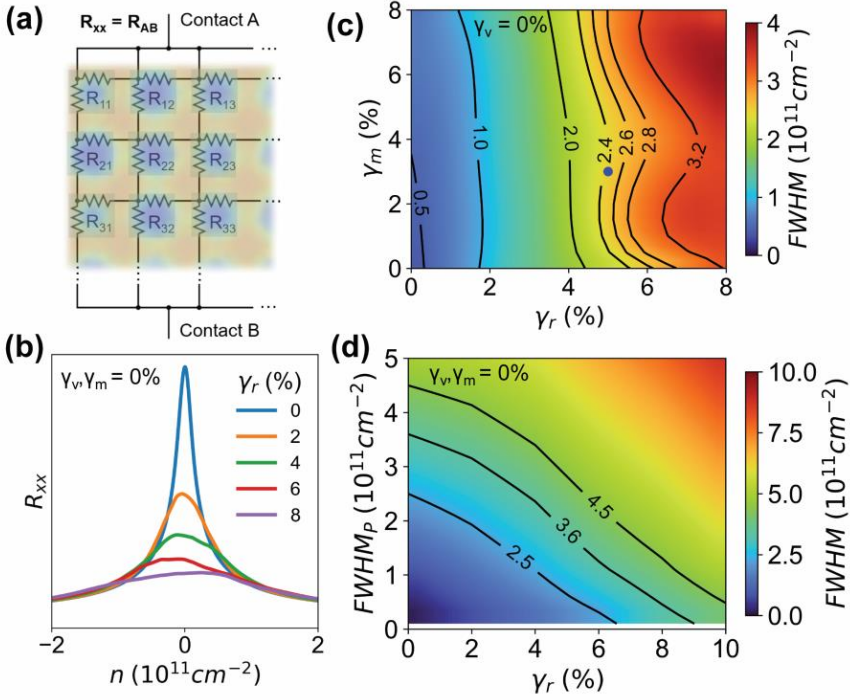

**Figure 4. Resistor network model to estimate disorder.** (a) Schematic of the resistor network used to simulate disorder. (b) Calculated $R_{xx}$ versus $n$ for varying $\gamma_r$. (c) FWHM of the primary Dirac peak as a function of $\gamma_r$ and $\gamma_m$. Blue dot represents the estimated disorder for our experimental superlattice device. (d) FWHM of the primary Dirac peak versus $\gamma_r$ and the FWHM$_P$ of pristine graphene.

# Supplementary Information for

## Understanding Disorder in Monolayer Graphene Devices with Gate-Defined Superlattices

Vinay Kammarchedu,[1,2,a,b] Derrick Butler,[1,2,4,a] Asmaul Smitha Rashid,[1,2] Aida Ebrahimi,[1,2,3,b] and Morteza Kayyalha[1,2,b]

[1]*Department of Electrical Engineering,*
[2]*Materials Research Institute,*
[3]*Department of Biomedical Engineering*
*The Pennsylvania State University, University Park, Pennsylvania 16802, United States*
[4]*Current address: National Institute of Standards and Technology, Gaithersburg, MD 20899, United States of America*

[a] V. Kammarchedu and D. Butler contributed equally to this work.
[b] **Author to whom correspondence should be addressed:** vinay@psu.edu, mzk463@psu.edu, sue66@psu.edu

# S1. Device fabrication and topographic analysis

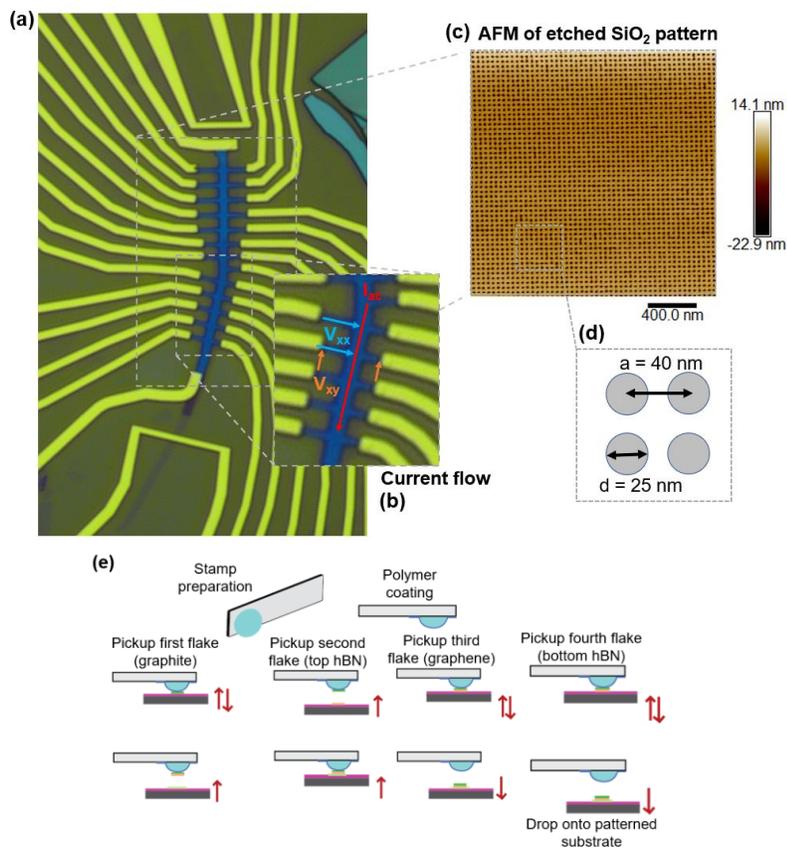

**Figure S1**. (a) Optical image of the fabricated Hall-bar structure. (b) Schematic of the current flow and resistance measurement contacts. (c) Atomic force microscope (AFM) image of the nanoholes prior to the heterostructure transfer. (d) Schematic of the designed superlattice. (e) Schematics of the heterostructure assembly process.

## S2. Analysis of three types of disorder in the patterned superlattice.

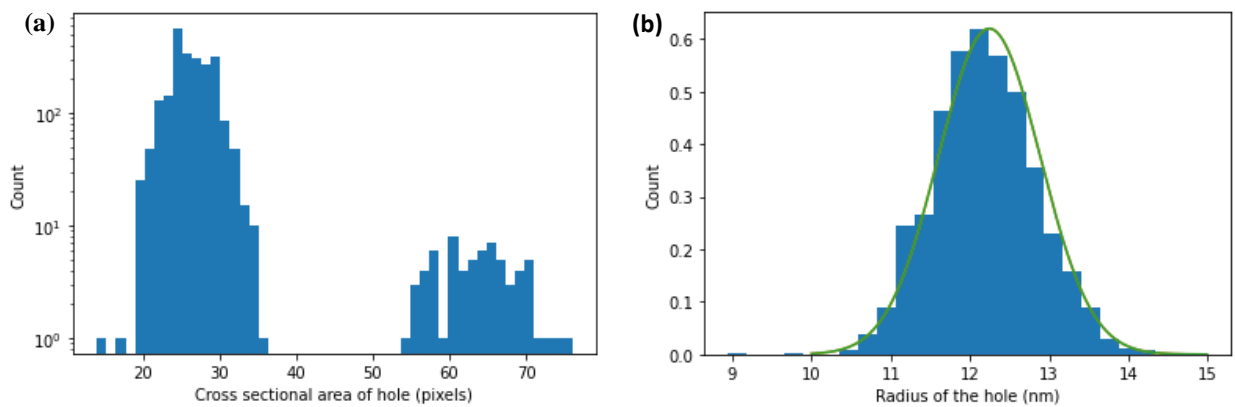

**Figure S2**. (a) Histogram of calculated hole cross sectional areas using image processing. The cluster at higher area (~ 60 pixels) is due to fusion of neighbouring holes. (b) Histogram of calculated hole radii. The distribution is normal with a standard deviation of 5%.

## S3. Intrinsic carrier density calculation.

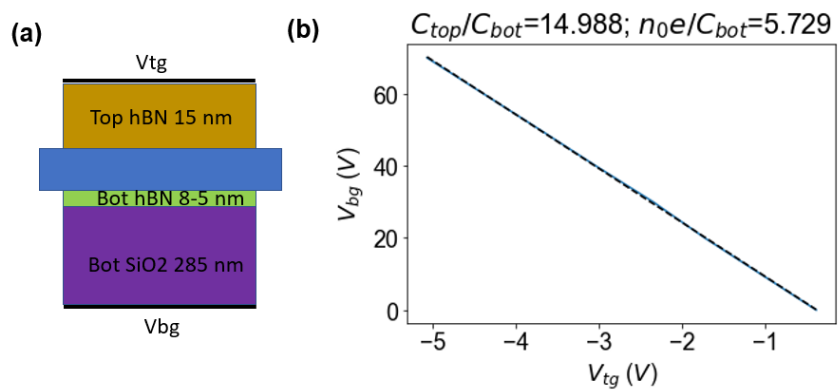

**Figure S3**. (a) Cross sectional schematic of the heterostructure. (b) Trajectory of the main Dirac point as a function of back- and top-gate voltages.

## S4. Additional transport data at zero magnetic field.

**Moiré peaks from hBN/graphene misalignment:** We also observe two satellite moiré peaks arising from unintentional alignment of graphene and hBN (twist angle of ~ 1.66° [1]) at carrier concentration of $|n| \approx 75 \times 10^{11}$ cm$^{-2}$. We note that this SL peak occurs at more than 10 times higher carrier concentration compared to the expected location of the patterned SL peaks, hence we expect little interference between these peaks. We also observe that satellite peaks at $|n| = 75 \times 10^{11}$ cm$^{-2}$ do not show any back gate-dependence, further suggesting that they are due to the moiré SL formed by the hBN-graphene misalignment.

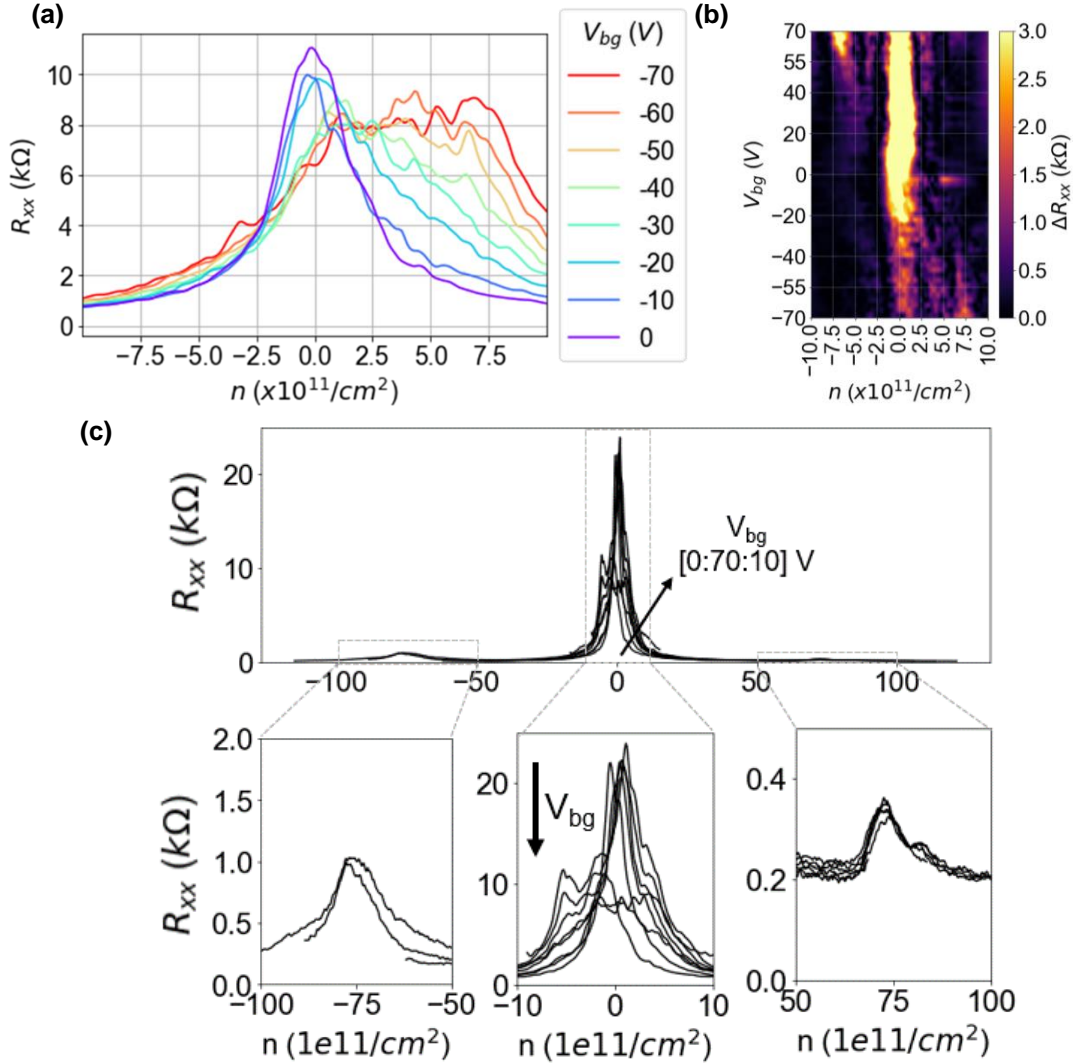

**Figure S4**. (a) Longitudinal resistance $R_{xx}$ as a function of carrier density $n$ at $B_z = 0$ for various negative back-gate voltages $V_{bg}$'s. (b) Baseline subtracted resistance map showing distinct peaks for $V_{bg} \geq 40$ V. (c) $R_{xx}$ as a function of $n$ at $B_z = 0$ and varying $V_{bg}$ for a larger range of carrier concentration. Left and right insets are zoomed in around the moiré mini-Dirac points, arising from graphene and hBN misalignment. Middle inset is zoomed in near the primary Dirac point.

## S5. Quantum transport simulations.

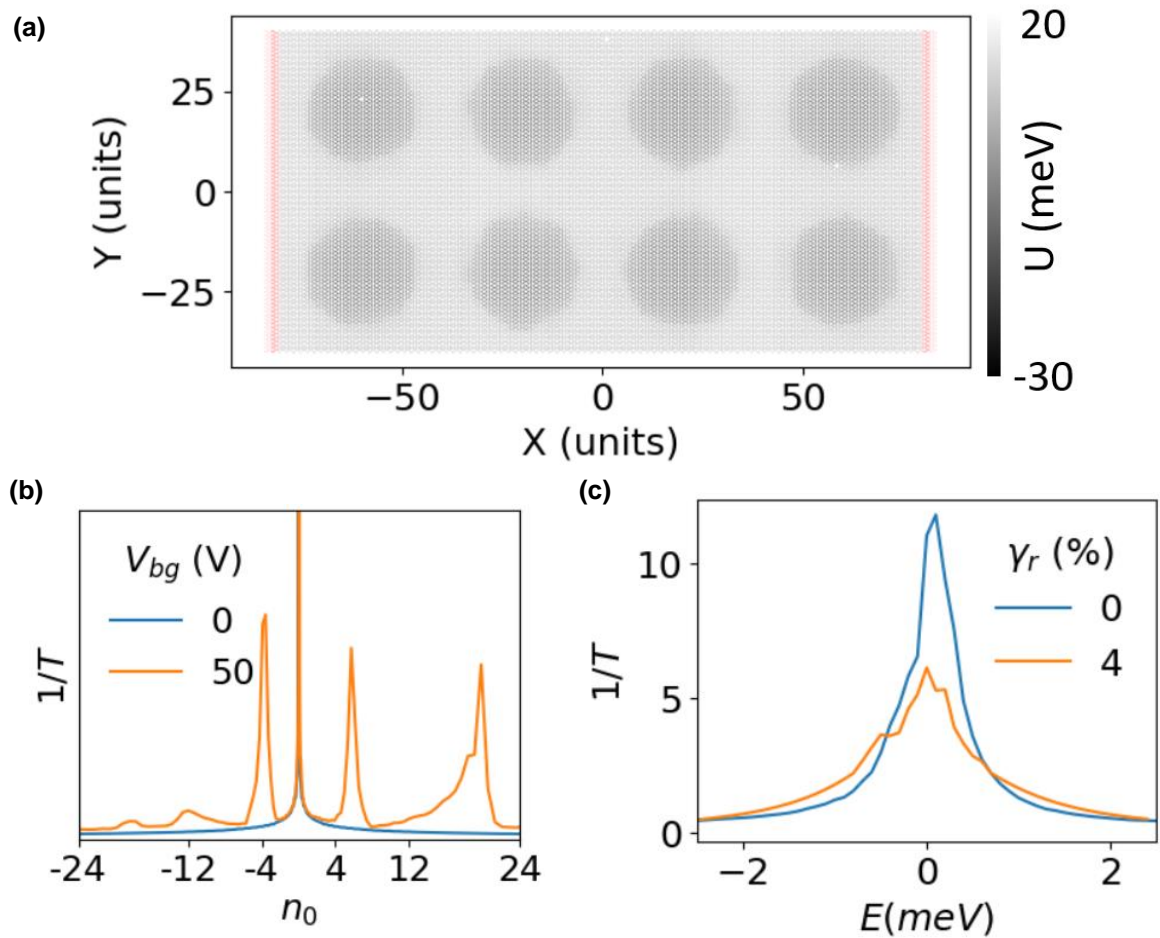

**Figure S5**. (a) Schematically shows a zoomed in version of the array of nanoholes (actual simulation is for a nanohole array of ~ 0.5 µm²) and graphene simulated along with the superlattice potential (U). Leads are present at the edge and are highlighted in red. (b) Plot of inverse of transmission (1/T, qualitatively equivalent to resistivity) vs. expected carrier concentration of first superlattice peak (4n$_0$) for $V_{bg} = 0$ and 50 V. (c) Plot of inverse of transmission $1/T$ versus energy for $\gamma_r = 0$ and 4%.

**S6. Magneto-transport results.**

We perform quantum Hall measurements at temperature $T = 12$ mK. Because of the SL potential, we expect to observe Hofstadter's fractal spectrum. This is obtained by the Diophantine equation:

$$\frac{n}{n_0} = t\left(\frac{\Phi}{\Phi_0}\right) + s, \qquad \text{Eq. S1}$$

where $(s, t)$ are two integers. We plot the longitudinal ($\sigma_{xx}$) and Hall ($\sigma_{xy}$) conductivity as a function of $n$ for $V_{bg}= 0$ V and $B_z = 8$ T as shown in Fig. S6a. We observe integer quantum Hall plateaus with filling factors of 2, 6, 10, etc. The filling factors show the typical sequence for monolayer graphene devices. We also plot longitudinal conductivity $\sigma_{xx}$ as a function of $n$ for various $V_{bg}$'s, showing the auxiliary peak that appears with increasing $V_{bg}$ at $B_z = 8$ T as shown in Figs. S6b and c. We construct a fan diagram by fixing the back-gate voltage at 70 V and sweeping the magnetic field, $B_z$, perpendicular to the device. Figure S6d shows $\sigma_{xx}$ as a function of $B_z$ and carrier concentration $n$. We observe a superposition of many fans, some of which we believe are due to the patterned SL effects. We identify eighteen different minima in the resistance from the fan diagram of Fig. S6d as marked by dashed lines. Theoretically, at lower gate voltages only peaks corresponding to Landau gaps should be observed, whereas at higher gate voltages, Hofstadter mini gaps should also be seen in the longitudinal conductance, $\sigma_{xx}$, data. These peaks are different from the Landau gap peaks with $s = 0$, i.e. a peak in $\sigma_{xx}$ corresponds to a level shift in the Hall conductance, $\sigma_{xy}$ (see Figs. S6a and c). However, the peaks that we associate to the SL have non-zero $s$, i.e. a peak in $\sigma_{xx}$ is seen while $\sigma_{xy}$ remains unchanged [2]. By extrapolating the dashed lines, we find an x-axis crossing at $B_z = 0$ T corresponding to the neutrality point of graphene. Furthermore, we observe x-axis crossings at $n = -5, -2.5, 2.5, 5 \times 10^{11}$ cm$^{-2}$ (white lines). The first crossing at $n = 2.5 \times 10^{11}$ cm$^{-2}$ is consistent with the mini peaks in Fig. 1c (in MS) and the calculated value of $4n_0$. Since the carrier concentration of the first peak is comparable to FWHM of the Dirac peak in our device, we observe clearer SL effects at higher back gate voltages and $B_z$ compared to other reports [2].

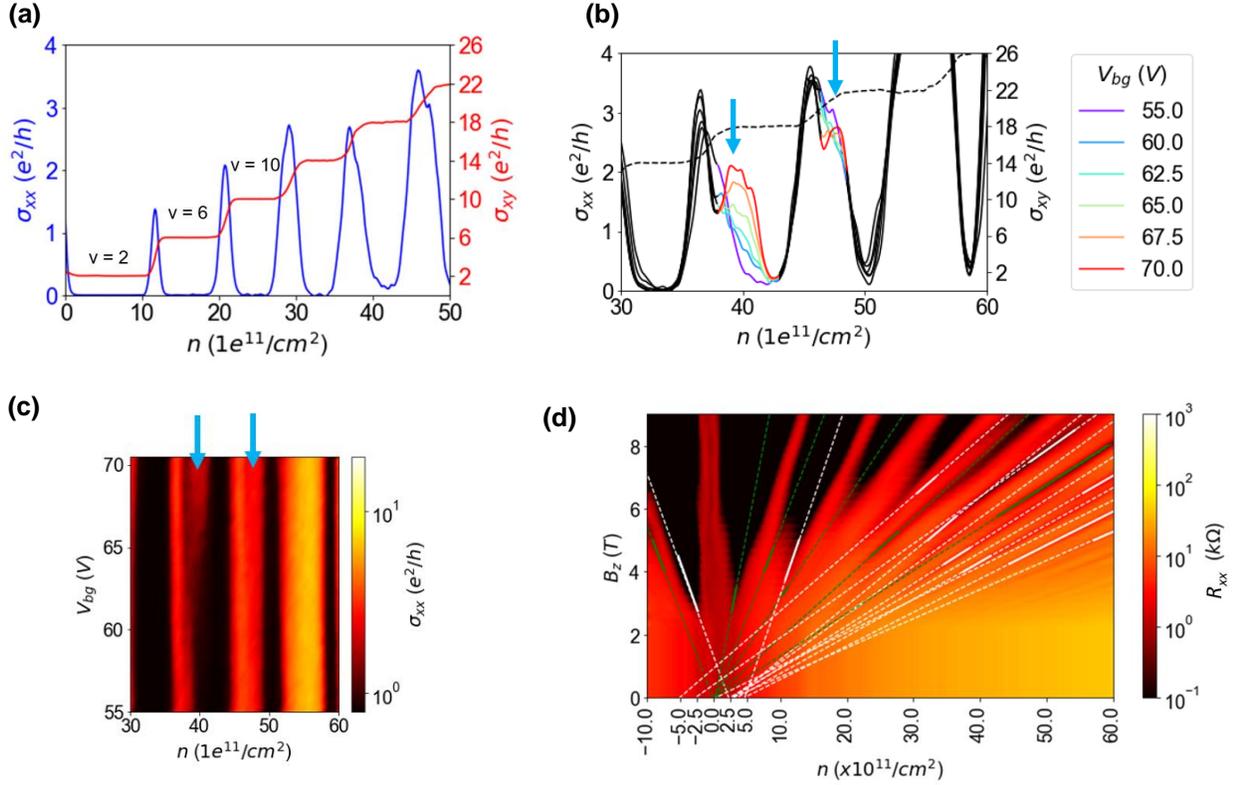

**Figure S6.** (a) Longitudinal ($\sigma_{xx}$) and Hall ($\sigma_{xy}$) conductivity as a function of $n$ for $V_{bg}= 0$ V and $B_z = 8$ T, showing the quantum Hall plateaus for filling factors 2, 6, 10, etc. (b) Longitudinal conductivity $\sigma_{xx}$ as a function of $n$ for various $V_{bg}$'s, showing the auxiliary peak that appears with increasing $V_{bg}$ at $B_z = 8$ T. (c) Colormap of $\sigma_{xx}$ as a function of $V_{bg}$ and $n$ at $B_z = 8$ T. An auxiliary peak (indicated by blue arrows) appears with increasing $V_{bg}$. (d) The device fan diagram at $V_{bg} = 70$ V. Extrapolating the minima leads to the identification of their origins at $B_z = 0$ T. The identified origins are attributed to the superlattice peaks around $n = -5, -2.5, 0, 2.5, 5 \times 10^{11}$ cm$^{-2}$.

## S7. Modeling disorder of a microscopic graphene without superlattice effects

In order to model the intrinsic disorder of a microscopic pristine device, we use an analytical equation (Eq. S1) to obtain the resistance (units: kOhm) as a function of the carrier concentration $n$ (units: $10^{11}$ cm$^{-2}$). The equation parameters are selected such that the FWHM of this function is ~ $0.25 \times 10^{11}$ cm$^{-2}$ as shown in Fig. S7.

$$R(n) = \sum_{i=-2}^{i=+2} (|n + (i \times 0.01)| + 0.1)^{-1} \qquad \text{Eq. S1}$$

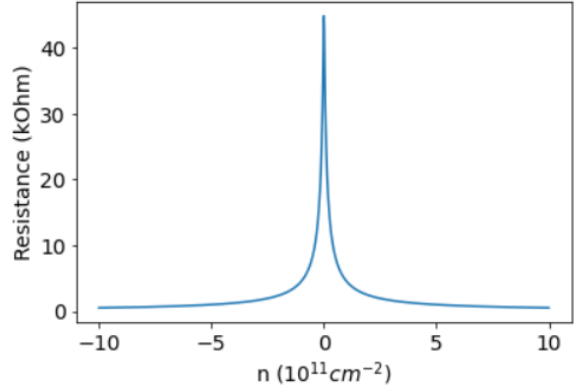

**Figure S7**. Plot of the resistance of a microscopic graphene device modeled with an intrinsic disorder leading to a full width half maximum of $0.25 \times 10^{11}$ cm$^{-2}$.

## S8. Disorder in triangular superlattice.

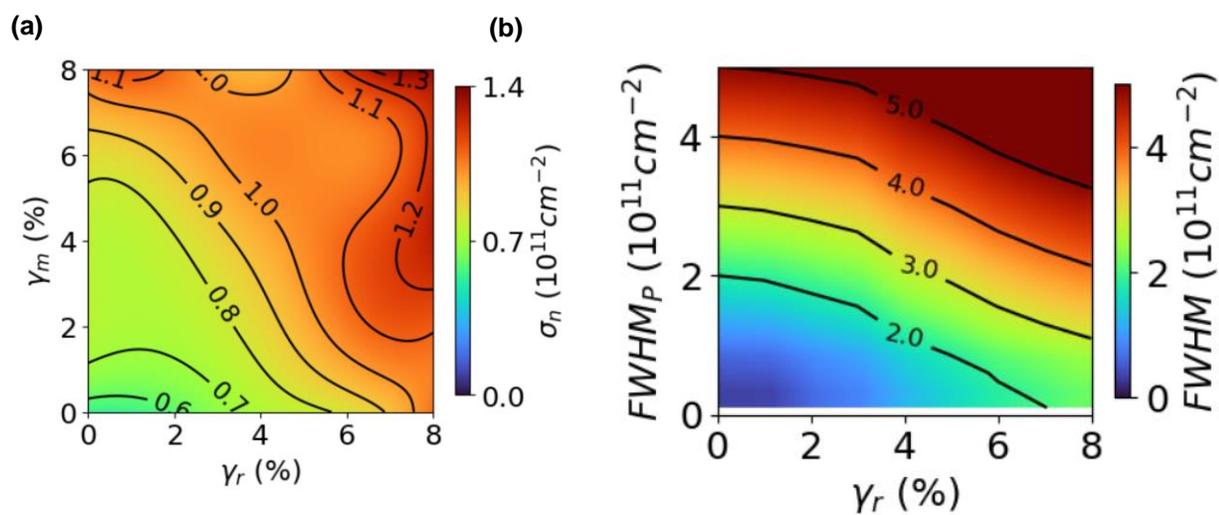

**Figure S8**. (a) The standard deviation $\sigma_n$ as a function of $\gamma_r$ and $\gamma_m$ for $\gamma_v = 0\%$. (b) The calculated FWHM of $R_{xx}$ versus $\gamma_r$ and $FWHM_p$ for $\gamma_v = \gamma_m = 0\%$.